\documentclass[prl,twocolumn,prl]{revtex4-1}

\usepackage{amsmath}
\usepackage{amssymb}
\usepackage{xspace}
\usepackage{graphicx}
\usepackage{grffile}
\usepackage{nicefrac}
\usepackage{color}

\newcommand{\bw}[1]{\raisebox{1.5ex}[-1.5ex]{#1}}
\begin{document}

\title{Unconventional electron states in $\delta$-doped SmTiO$_3$}

\author{Frank Lechermann}
\affiliation{I. Institut f{\"u}r Theoretische Physik, Universit{\"a}t Hamburg, 
D-20355 Hamburg, Germany}
\affiliation{Institut f\"ur Keramische Hochleistungswerkstoffe, Technische
Universit\"at Hamburg-Harburg, D-21073 Hamburg, Germany}


\begin{abstract}
The Mott-insulating distorted perovskite SmTiO$_3$, doped with a single SrO layer in 
a quantum-well architecture is studied by the combination of density functional 
theory with dynamical mean-field theory.
A rich correlated electronic structure in line with recent experimental investigations 
is revealed by the given realistic many-body approach to a large-unit-cell oxide
heterostructure.
Coexistence of conducting and Mott-insulating TiO$_2$ layers prone to magnetic order
gives rise to multi-orbital electronic transport beyond standard Fermi-liquid theory. 
First hints towards a pseudogap opening due to electron-electron scattering within a 
background of ferromagnetic and antiferromagnetic fluctuations are detected.  
\end{abstract}

\maketitle

Doped Mott insulators pose a challenging condensed matter problem
(see e.g.~\cite{ima98} for a review). At stoichiometry, simple correlated metals show 
renormalized Landau-like quasiparticles, while charge-gapped Mott (including charge-transfer) 
insulators often reveal long-range order at low temperature with again a Landau-like order 
parameter. On the contrary, prominent materials such as e.g. high-T$_{\rm c}$ cuprates, 
double-exchange driven manganites or the correlated-spin-orbit iridate family prove that 
doping a Mott-insulating state can give rise to novel intricate phases, often beyond the 
Landau paradigm.

The in-depth experimental and theoretical analysis of the effect of random 
doping in bulk systems is usually hindered by the impact of disorder on 
introduced charges and local structural relaxations. This renders the definition of relevant 
length scales, e.g. screening distances, difficult. Due to the complexity of the problem, 
many theoreties of doped correlated materials, especially on the model-Hamiltonian 
level, neglect details of the local-chemistry aspect. But this may be insufficient
to elucidate the subtle energy-scale balancing of strongly correlated electrons systems 
prone to long-range order.

Two developments are eligible to shed new light on this longstanding problem. First the
rising field of oxide heterostructures allows experimentalists to introduce well-defined
doping layers in correlated materials~\cite{eck95,ste13}. Thereby the problem of disorder and 
ambiguities in identifying unique length scales are removed. Second the combination of 
first-principles density functional theory (DFT) with dynamical mean-field theory (DMFT) 
accounts for the interplay of bandstructure features and many-body effects 
beyond the realm of static-correlation approaches~\cite{ani97,lic98}. Allying these 
progresses by addressing a doped-Mott-insulator heterostructure via DFT+DMFT is thus 
proper to reveal new insight into a hallmark challenge of interacting electrons.

The distorted perovskite SmTiO$_3$ as a member of the $R$TiO$_3$ ($R$: rare-earth element) series 
with formal Ti$^{3+}$-$3d(t_{2g}^1)$ configuration is Mott-insulating at stoichiometry. It displays 
antiferromagnetic (AFM) ordering below $T_{\rm N}=45$K. Notably, in the given $3d^1$ titanate 
series the compound is just at the border of a quantum-critical transformation from 
antiferromagnetic to ferromagnetic (FM) order~\cite{kom07}. 
Recent experimental work focusing on $\delta$-doping SmTiO$_3$
by a single SrO layer exposed non-Fermi-liquid (NFL) character, with a subtle crossover 
to still intriguing transport behavior when adding further doping layers~\cite{jac14,mik15,mik16}.
\begin{figure*}[t]
\parbox[b]{4.7cm}{(a)\hspace*{-0.4cm}\includegraphics*[height=8.25cm]{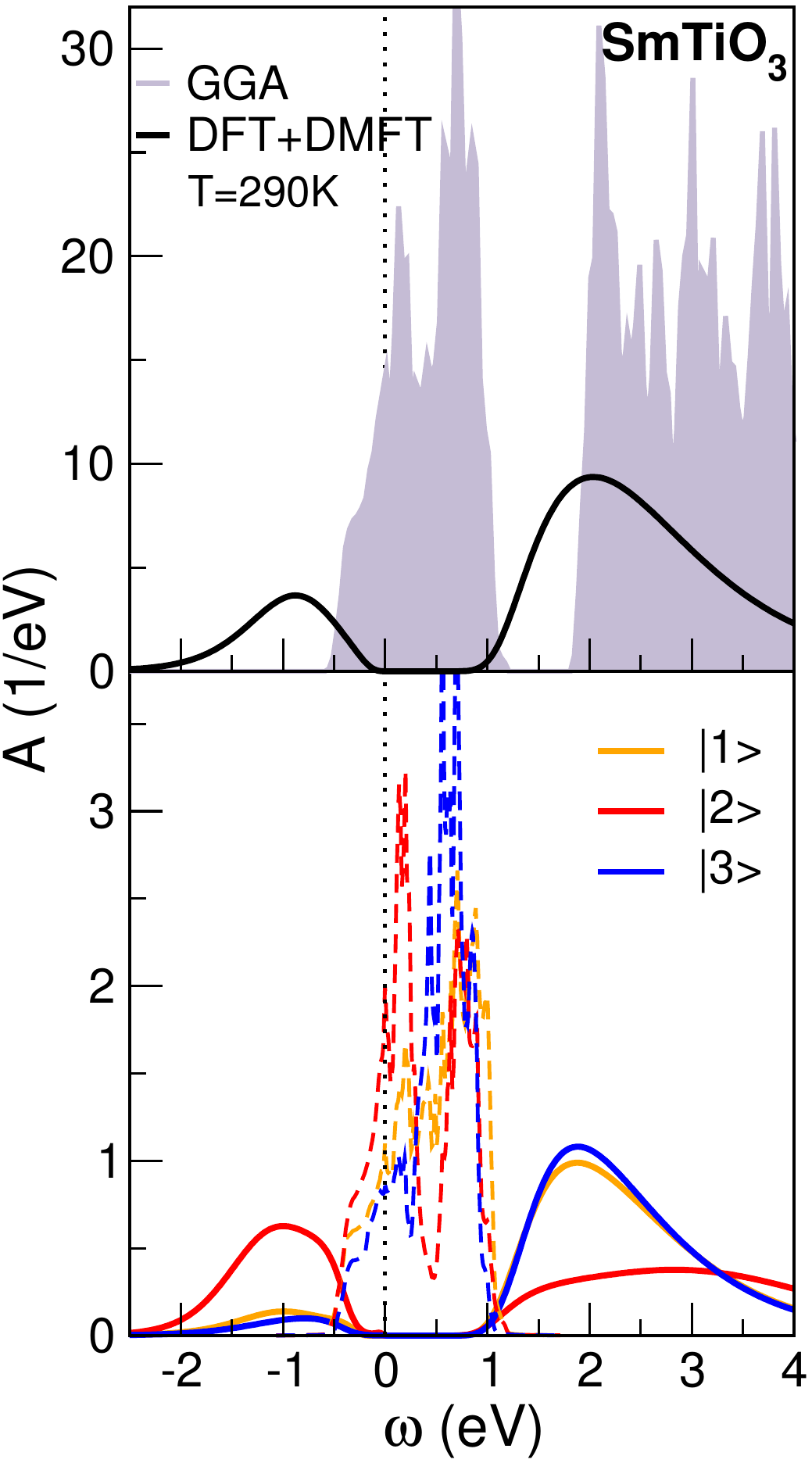}}
\parbox[b]{3.7cm}{\includegraphics*[height=8.25cm]{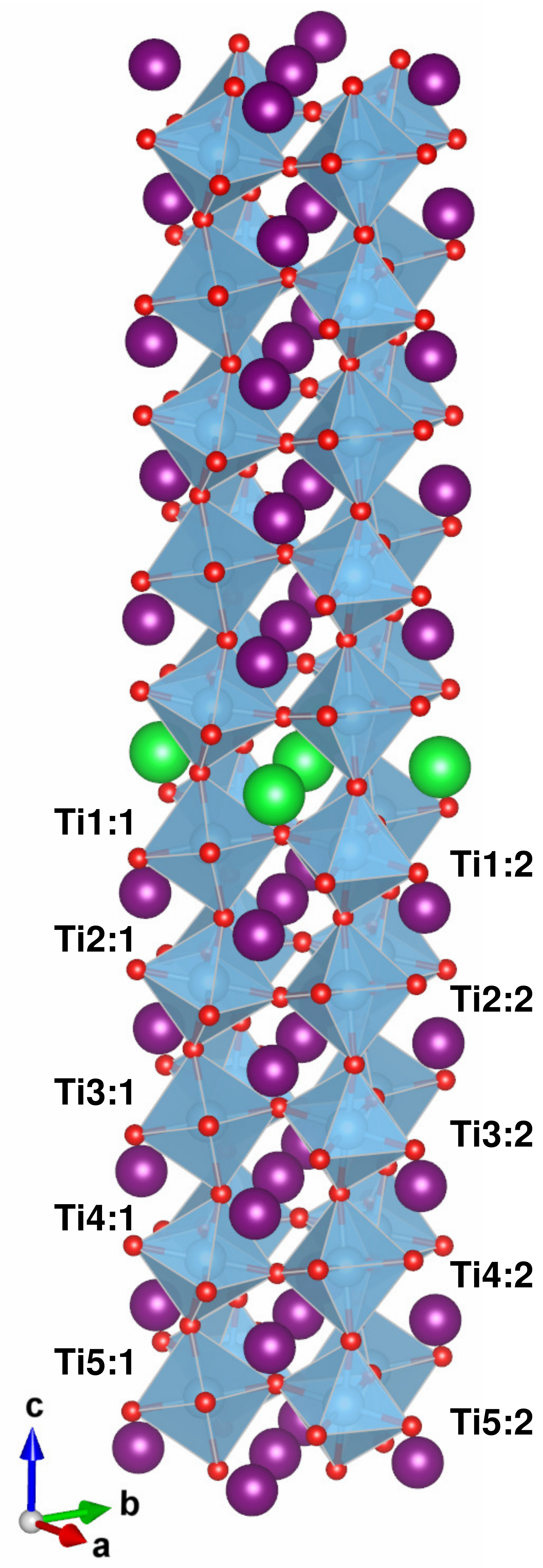}\\[0.1cm]
(b)\hspace*{-0.2cm}\includegraphics*[width=3.5cm]{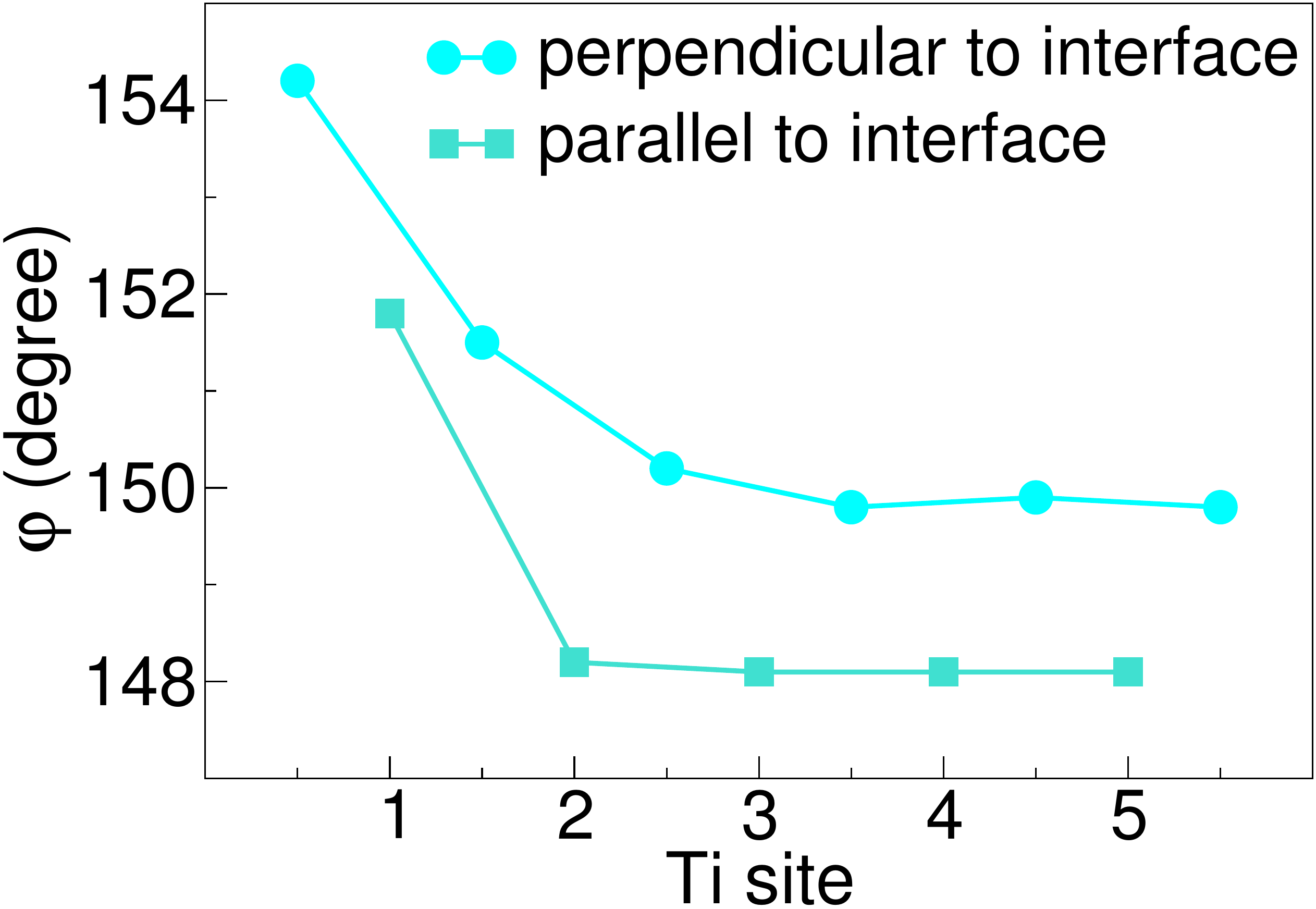}}
\parbox[b]{9cm}{
(c)\hspace*{-0.3cm}\includegraphics*[width=8.7cm]{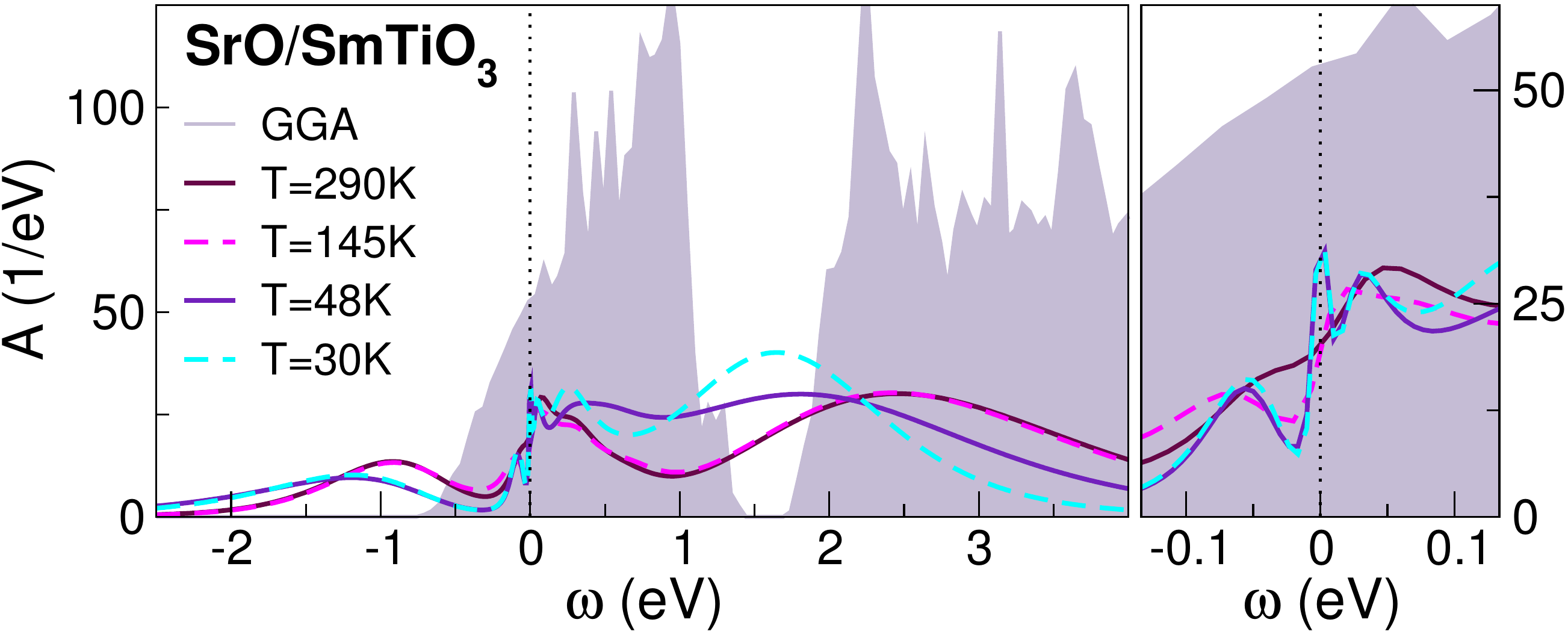}\\[0.2cm]
(d)\hspace*{-0.1cm}\includegraphics*[width=8.7cm]{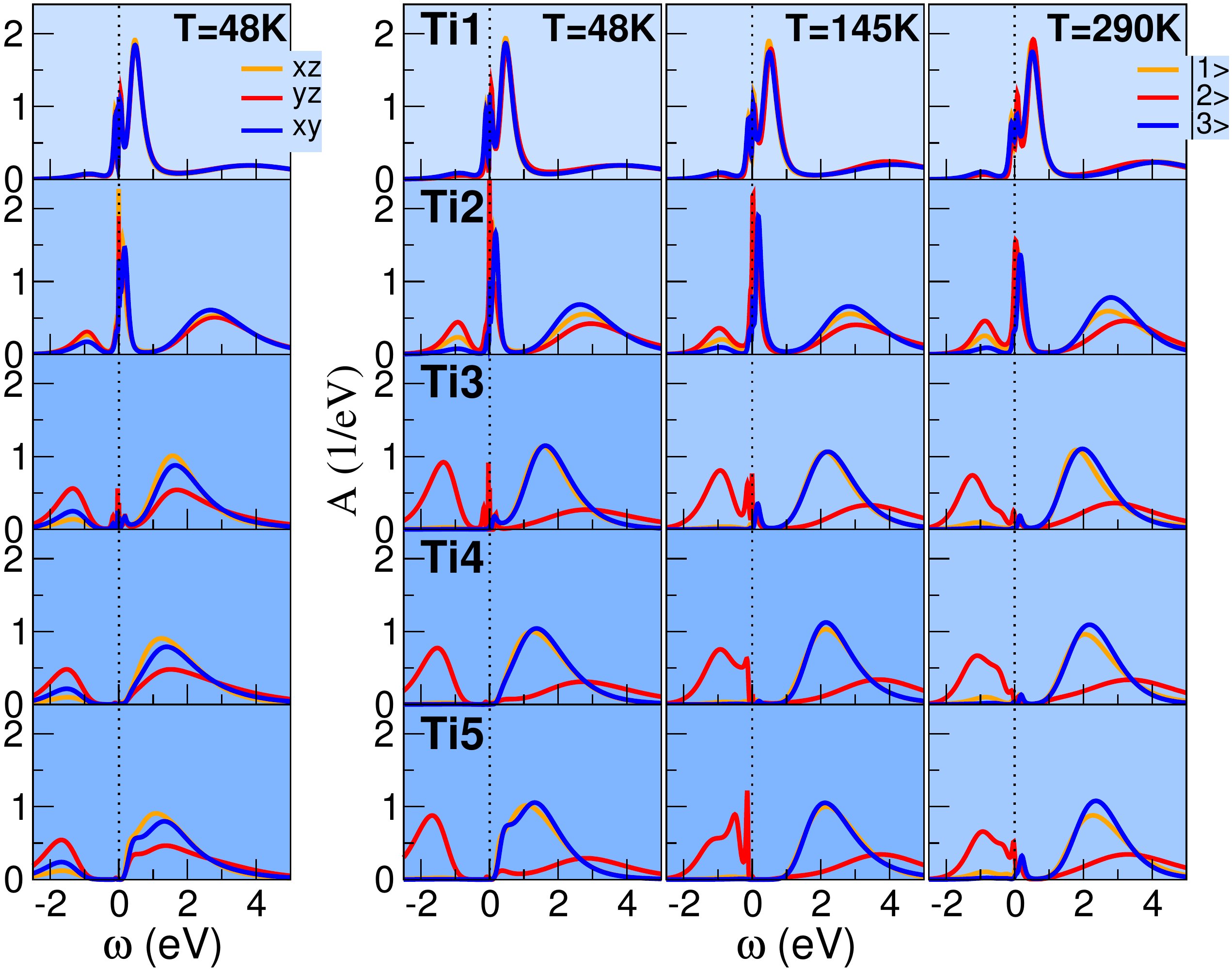}}
\caption{(color online) DFT+DMFT results for paramagnetic undoped/$\delta$-doped 
SmTiO$_3$. (a) Correlated electronic structure of stoichiometric SmTiO$_3$. (b) Top: supercell 
of the $\delta$-doped compound: Sm (violet), Sr (green), Ti (lightblue), O (small red), bottom: 
Ti-O-Ti bond angles. (c,d) Spectral function of the $\delta$-doped system for (c) the total 
system and (d) Ti-resolved, left: for a single temperature in the conventional cubic $t_{2g}$
basis and right: over a wider $T$ range in the symmetry-adapted Ti-dependent effective $t_{2g}$ 
basis.}\label{fig1:para}
\end{figure*}

In this work, a realistic many-body approach is employed to resolve the multi-orbital 
correlated electronic structure of $\delta$-doped SmTiO$_3$. We reveal a coexistence between 
itinerant and Mott-insulating real-space regions associated with different orbital 
polarizations. Non-Fermi-liquid behavior originates from the internal boundaries.
Eventually, the scattering of itinerant carriers with spin fluctuations near the designed
AFM-FM crossover is responsible for a pseudogap fingerprint, giving reason for the realistic 
NFL regime. These findings pave the way for theoretical investigations of oxide interfaces 
conducted by materials-design approaches beyond the possibilities of static mean-field studies.

Charge self-consistent DFT+DMFT~\cite{sav01,pou07,gri12} is used to access the many-body
correlated electronic structure, using a correlated subspace composed of effective (i.e. 
Wannier-like)~\cite{ama08,ani05,aic09,hau10} Ti $3d(t_{2g})$ orbitals 
$w(t_{2g})$~\cite{supp}. Local Coulomb interactions in Slater-Kanamori 
form are parametrized by a Hubbard $U=5\,$eV and a Hund's coupling 
$J_{\rm H}=0.64\,$eV~\cite{pav04}. The multi single-site DMFT impurity problems~\cite{pot99} 
are solved by the continuous-time quantum Monte Carlo scheme~\cite{rub05,wer06,par15,set16}.

Let us focus first on stoichiometric SmTiO$_3$ (cf. Fig.~\ref{fig1:para}a).
Besides the lattice parameters, characteristic for the GdFeO$_3$-type distorted-perosvskite 
structure (space group $Pbnm$) are the Ti-O(1,2)-Ti bond angles, whereby O1(2) is 
the apical(basal in-plane) oxygen position with respect to the $c$-axis~\cite{kom07}. 
In bulk SmTiO$_3$, these angles read $\varphi_1,\varphi_2=146^\circ, 147^\circ$. Based on the
experimental crystal data~\cite{kom07}, the Ti($t_{2g}$) states form an isolated low-energy 
metallic band manifold of width $W~\sim 1.55$eV in DFT. A small orbital polarization towards an 
nearly isotropic effective $t_{2g}$ state
$|2\rangle=0.58|xz\rangle +0.53|yz\rangle +0.62|xy\rangle$ is detected. The remaining two effective
$t_{2g}$ orbitals are given by $|1\rangle=0.76|xz\rangle -0.63|yz\rangle -0.17|xy\rangle$ and
$|3\rangle=0.30|xz\rangle +0.57|yz\rangle -0.78|xy\rangle$. In line with experiment, 
strong electron correlations drive the material paramagnetic (PM) Mott insulating by
effectively localizing a single $t_{2g}$ electron on the Ti site. Furthermore, 
as observed in theoretical assessments of other Mott-insulating $3d^1$ 
titanates~\cite{pav04,pav05,lec15}, a substantial orbital polarization, here towards 
state $|2\rangle$, occurs. The orbital occupation reads $(n_1, n_2, n_3)=(0.15, 0.75,0.10)$. 
If we define the charge gap $\Delta_g$ by spectral weight $<10^{-4}\,{\rm eV}^{-1}$, a 
value $\Delta_g=0.55$eV is obtained, in good agreement with the mid-infrared-absorption 
onset of 0.50eV~\cite{cra92}. Below its N\'eel temperature, bulk SmTiO$_3$ becomes an
G-type antiferromagnet in experiment. Calculations show~\cite{supp} that still various 
magnetic orderings are nearly degenerate in energy, in line with the system being on the 
verge to an AFM-FM transition.

Our 100-atom-unit-cell superlattice establishes $\delta$-doping of SmTiO$_3$ by
a single SrO layer~\cite{supp}. It incoporates five symmetry-inequivalent TiO$_2$ layers each 
with two lateral inequivalent Ti sites (see Fig.~\ref{fig1:para}b). 
The Mott insulator is doped with two holes, i.e. nominally 0.1 hole per Ti. In the experimental 
setting of Ref.~\cite{jac14,mik15} the original $c$-axis is {\sl parallel} to SrO and the original 
$a,b$-axes are inclined. To account for this fact approximatively, we bring the original lattice 
parameters~\cite{kom07} in the same directional form, but without lowering 
the $Pbnm$ symmetry and relax all atomic positions. At the doping layer the bond 
angles $\varphi_{1,2}$ are enhanced 
(see Fig.~\ref{fig1:para}b). There the system is structurally driven towards 
cubic SrTiO$_3$. As expected, beyond 3-4 layers the characteristic angles saturate to bulk-like 
values~\cite{che13}. Not-surprisingly, this saturation happens faster in terms of layers for 
the in-plane angle, since the out-of-plane $\varphi_1$ is stronger affected from a 
plane-parallel interface.

The PM many-body electronic structure upon $\delta$-doping is exhibited in 
Figs.~\ref{fig1:para}c,d. In line with experimental findings~\cite{jac14}, DFT+DMFT reveals 
metallicity, but with characteristics different from GGA. The total spectral 
function $A(\omega)$ shows strong band narrowing and transfer of spectral weight to 
Hubbard bands. These processes depend rather significantly on the temperature $T$.
Already for $T=145\,$K an obvious spectral reduction sets in at low-energy. This means that
the coherence scale for low-energy excitations is far more lower than in many other
correlated bulk systems. Comparison between $T=48\,{\rm K}, 30\,{\rm K}$ data shows that
the overall electronic structure finally settles well below the coherence scale, giving 
rise to a lower Hubbard band at around $-1.2$eV. Notably within a small $[-0.1$$,$$0.1]\,$eV 
energy window around the Fermi level, a three-peak quasiparticle (QP)-like structure emerges. 
It should not be confused with the conventional large-energy-range three-peak structure 
involving lower and upper Hubbard bands.
\begin{table}[b]
\begin{ruledtabular}
\begin{tabular}{c|rrrrr}
     orbital & Ti1  & Ti2  & Ti3  & Ti4  & Ti5  \\ \hline
$|1\rangle$  & 0.22 & 0.29 & 0.05 & 0.04 & 0.04 \\
$|2\rangle$  & 0.16 & 0.48 & 0.94 & 0.95 & 0.95 \\
$|3\rangle$  & 0.24 & 0.12 & 0.01 & 0.01 & 0.01 \\[0.1cm]
sum          & 0.62 & 0.89 & 1.00 & 1.00 & 1.00
\end{tabular}
\end{ruledtabular}
\caption{Temperature-averaged effective Ti($t_{2g}$) occupations within each TiO$_2$
layer of $\delta$-doped SmTiO$_3$.}\label{tab:occ}
\end{table}

We did not encounter intra- or inter-layer charge-ordering instabilities. Such charge 
ordering could even not be meta-stabilized. Especially the in-plane Ti ions behave equivalent, 
thus no need for intra-layer differentiation in the discussion. However the correlated subspace 
of $w(t_{2g})$ orbitals becomes layer-Ti dependent. Still, the Wannier-like functions in the 
different layers group again in the bulk-established subclasses, and the notion of
 $w(t_{2g})=|1\rangle,|2\rangle,|3\rangle$ orbitals in each layer remains coherently 
applicable.

Different electronic phases are detected with distance to the SrO doping layer 
(see Fig.~\ref{fig1:para}d). While the nearest TiO$_2$ layer is orbital-balanced conducting, 
the layers beyond the second one become strongly orbital-polarized Mott insulating at low $T$. 
The second layer itself is metallic, however displays strong correlations with already 
substantial orbital $|2\rangle$ polarization. With raising temperature, the more 
distant layers partly also obtain metallic character, but in a very incoherent fashion without 
clear QP formation. The orbital occupations only weakly depend on $T$ but have strong layer 
dependence (see Tab.~\ref{tab:occ}). As in the bulk, layers 3-5 localize one electron in the 
Ti($t_{2g}$) shell, whereas the 2nd layer with about 0.9 electrons is in a doped-Mott state. The
first layer with 0.6 electrons appears as a renormalized metal. Hence between different TiO$_2$ 
layers, intricate metal-insulator transitions with strong orbital signature and delicate $T$
dependence below room temperature are revealed.
\begin{figure}[t]
\centering
(a)\hspace*{-0.4cm}\includegraphics*[height=3.9cm]{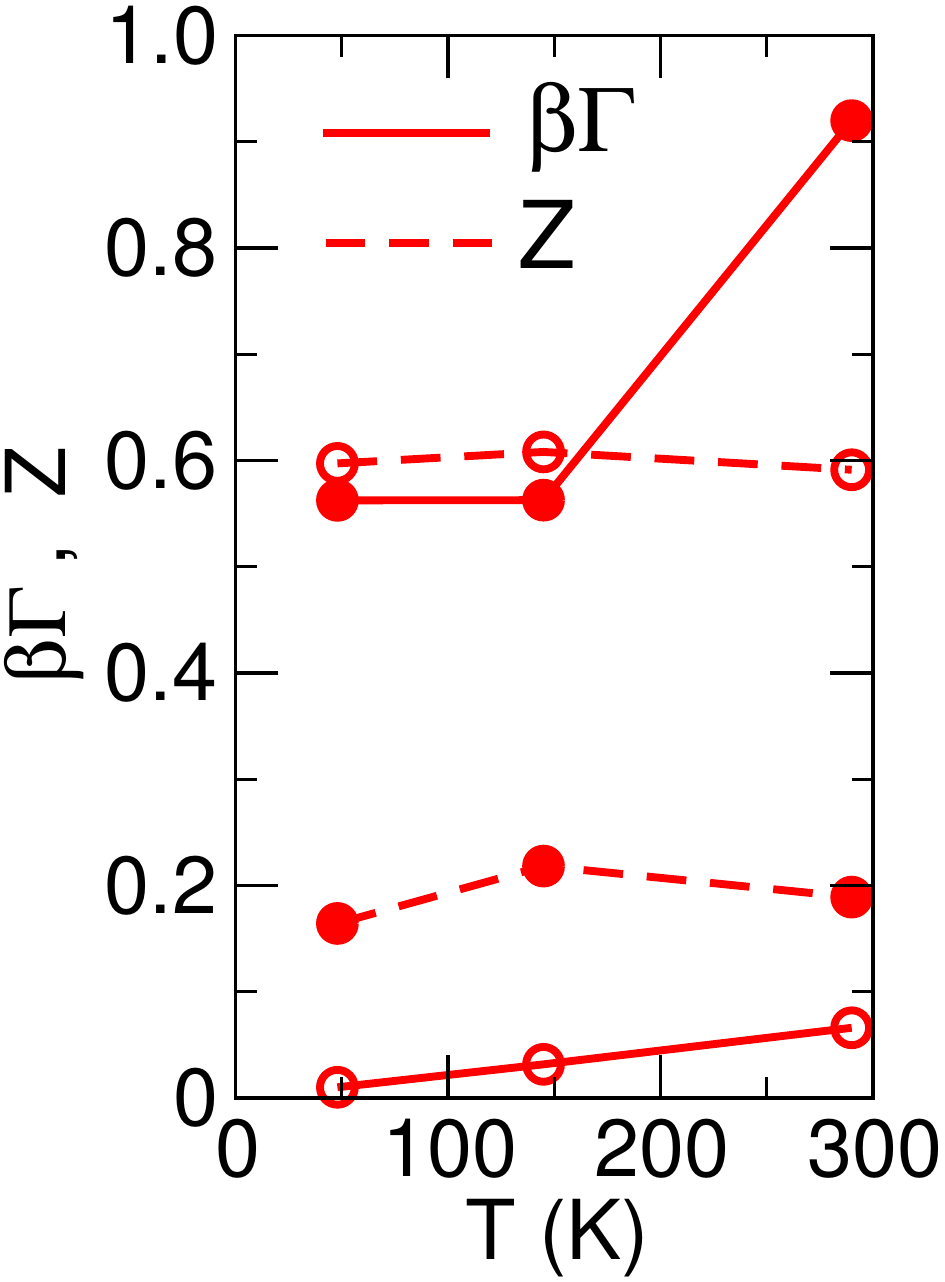}
\hspace*{-0.2cm}(b)\hspace*{-0.3cm}\includegraphics*[height=3.9cm]{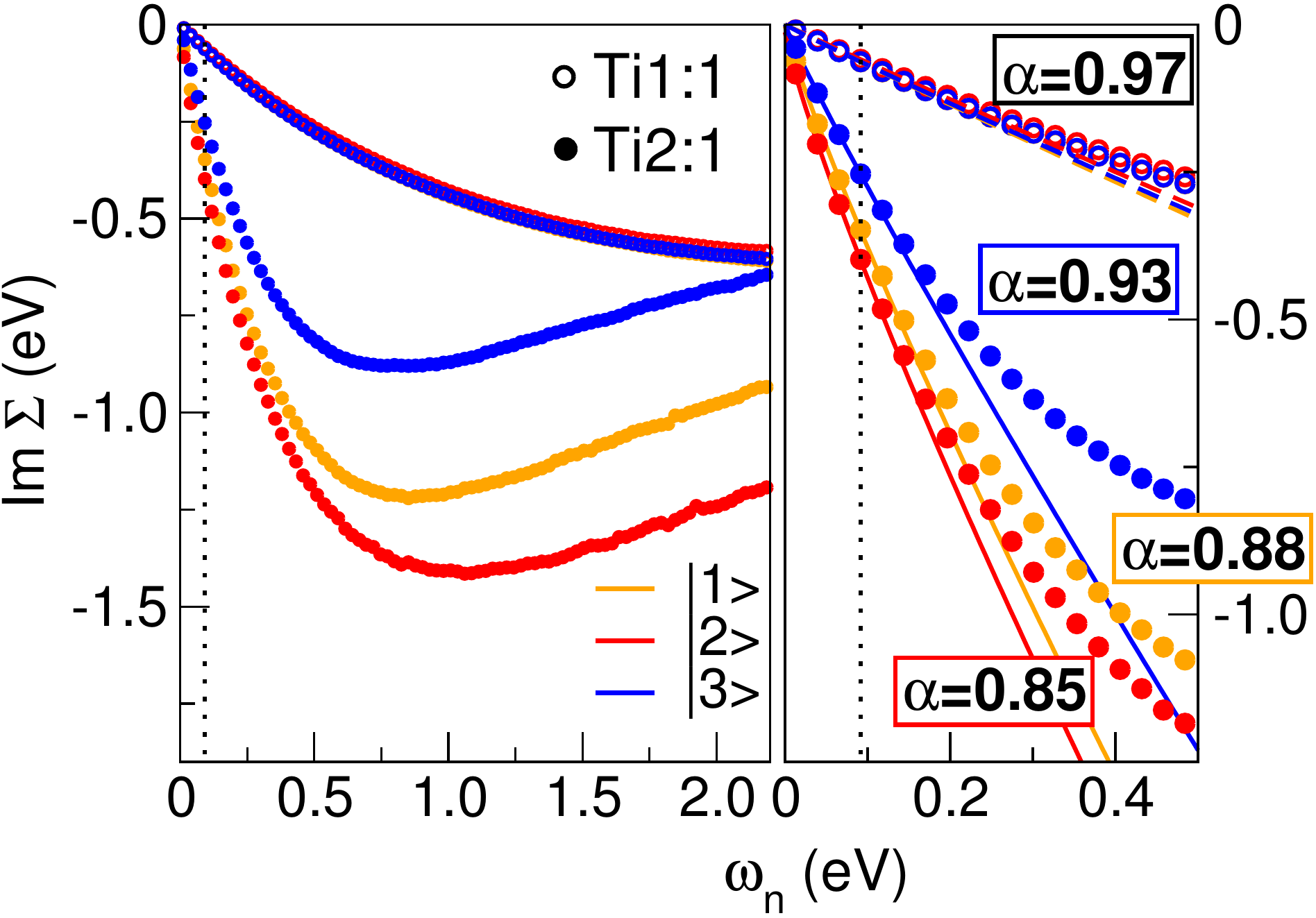}
\caption{(color online) Transport analysis for both conducting layers. 
(a) QP weight and scattering rate for the $|2\rangle$ state.
(b) Orbital-resolved imaginary part of the self-energy on the Matsubara axis
$\omega_n=(2n+1)\,\pi\,T$. 
Left: larger frequency range, right: low-frequency region with fitting 
functions ${\rm Im}\,\Sigma(\omega_n)=C_0+A\,\omega_n^\alpha$ (dashed/full lines).
Exponential-fitting cutoff $n_c$ is denoted by the dotted line~\cite{supp}.}
\label{fig2:self}
\end{figure}

For the rest of the paper, we refer to the nearest(next-nearest) TiO$_2$ plane with 
respect to the SrO doping plane as 'first(second) layer'. In order to assess the transport 
characteristics of these two conducting layers, the low-frequency behavior of the respective 
orbital-resolved self-energies $\Sigma(i\omega_n)$ is analyzed in 
Fig.~\ref{fig2:self}~\cite{supp}. {\sl Assuming} an overall Fermi-liquid regime, the QP weight 
$Z=(1-\frac{\partial\mbox{\scriptsize Im}\,\Sigma(i\omega_n)}{\partial\omega_n}|_{\omega_n\rightarrow 0^+})^{-1}=\frac{m^*_{\rm GGA}}{m^*}$ and the electron-electron scattering rate 
$\beta\Gamma=-\beta Z\,{\rm Im}\,\Sigma(i0^{+})$ is displayed, whereby $m^*$ denotes the
effective mass and $\beta=1/T$. Whereas the first layer indeed shows well-developed Fermi-liquid 
like scattering and moderate $Z_1\sim 0.6$, the second layer inherits strong scattering and a much
smaller formal $Z_2\sim 0.2$. To examine the quality of the Fermi-liquid character, an 
exponential-function fit is performed to the imaginary part of $\Sigma(i\omega_n)$, i.e.
${\rm Im}\,\Sigma(\omega_n)\stackrel{!}{=}C_0+A\,\omega_n^\alpha$~\cite{wer08}. 
An ideal exponent $\alpha=1$ marks a well-defined Fermi liquid with 
corresponding $T^2$-law for the resistivity. 
Here indeed Fermi-liquid-like values $\alpha_1=0.97$ and $C_0\rightarrow 0$ are extracted for 
the first layer, but for the dominant orbital $|2\rangle$ in the second layer an 
exponent $\alpha_2=0.85$ and finite intercept $C_0\sim -0.01\,$eV are obtained. 
Thus the second layer, mediating between Fermi-liquid and Mott-insulator, is put into a 
non-Fermi-liquid regime. This happens when the overall coherence scale is already reached, 
hence a conventional bad-metal picturing is not easily applicable. Note that in experiment, 
also a subtle NFL regime with $T^{5/3}$-law is measured for $\delta$-doped 
SmTiO$_3$~\cite{jac14}.

\begin{figure}[t]
\centering
\parbox[c]{2.1cm}{\hspace*{-0.4cm}\includegraphics*[width=2.1cm]{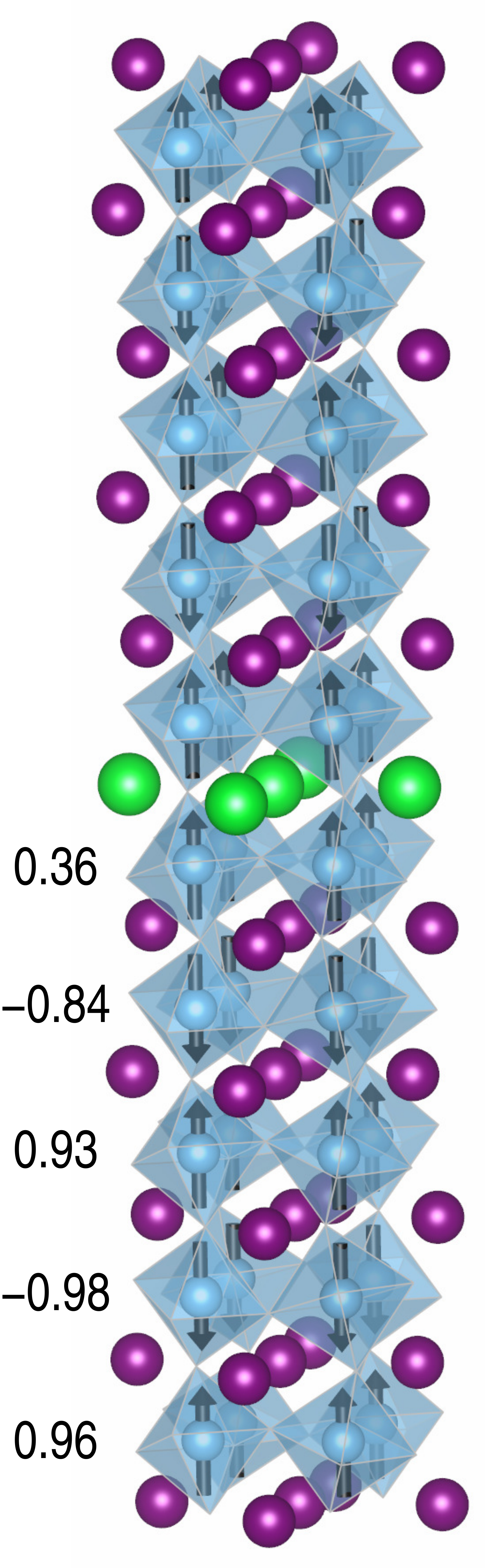}\\[0.1cm]
(a)}
\parbox[c]{6.1cm}{
(b)\hspace*{-0.4cm}\includegraphics*[width=6.1cm]{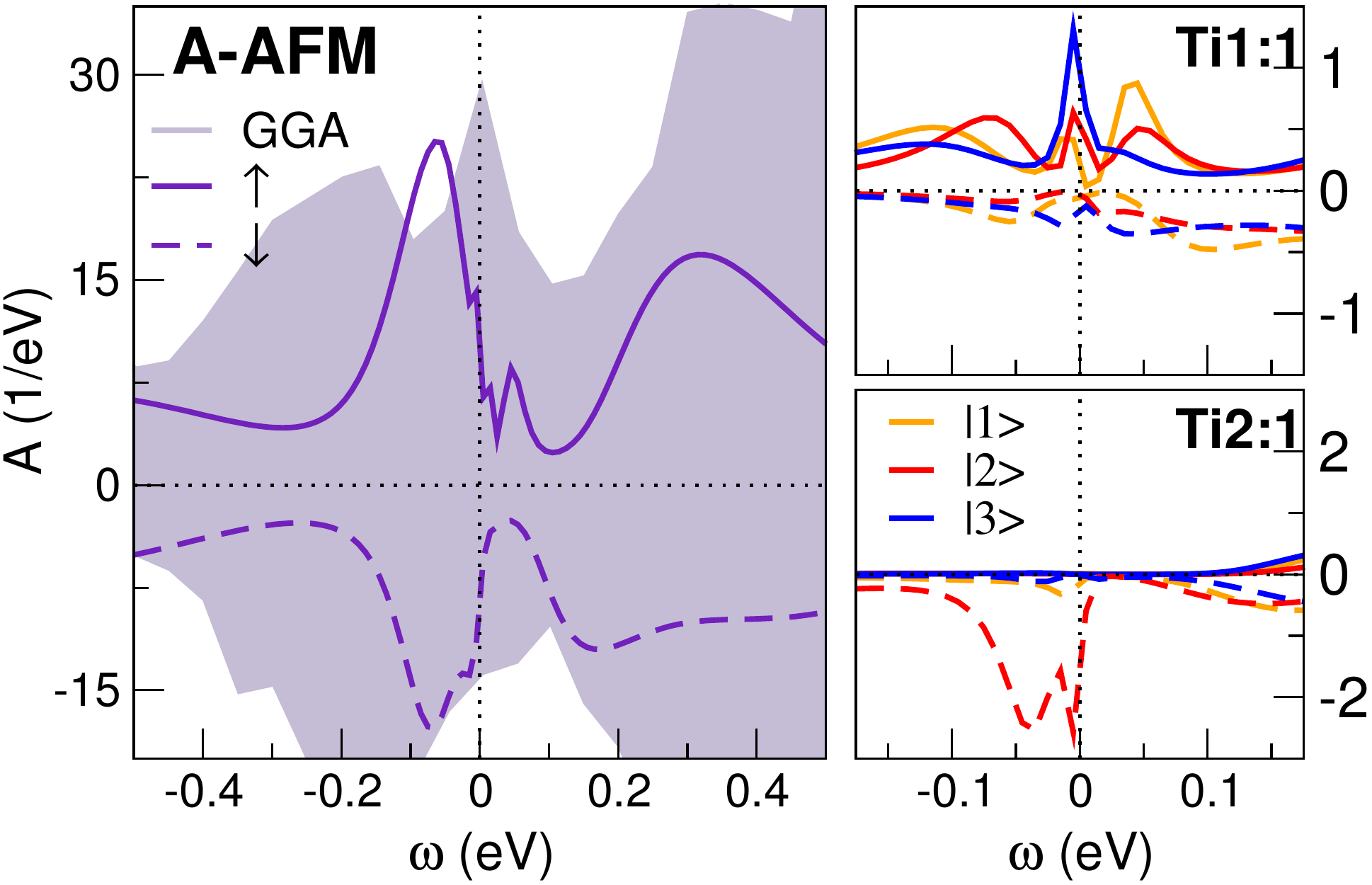}\\
(c)\hspace*{-0.4cm}\includegraphics*[width=6.1cm]{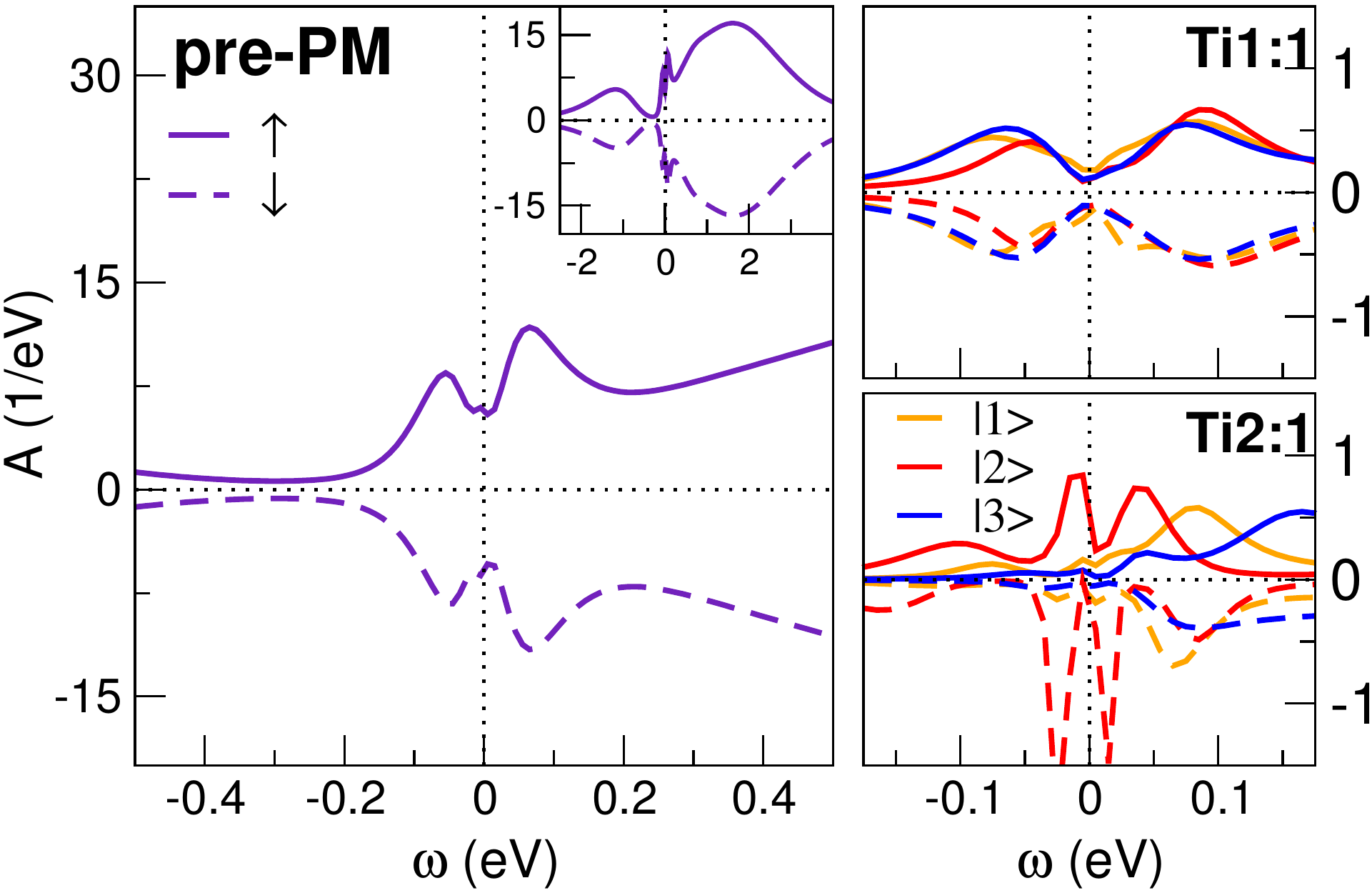}}
\caption{(color online) $\delta$-doped SmTiO$_3$ with broken spin symmetry ($T=48$K).
(a) A-AFM magnetic order, numbers provide local Ti magnetic moment in $\mu_{\rm B}$.
(b,c) Spectral function, left: total, right: local Ti for first and second layer
(similar data for Ti:2 sites). 
(b) A-type AFM phase. (c) pre-converged phase (after 20 DFT+DMFT steps), 
when starting the self-consistent calculation from the PM solution.}\label{fig3:pg}
\end{figure}

To shed further light onto the nature of the NFL behavior, possible broken-symmetry states are
taken into account. Albeit various initializing starting points are investigated, again 
(spin-broken-assisted) charge-ordering instabilities are not supported by the present 
theoretical schemes. On the other hand, A-type AFM ordering, i.e. intra-layer FM and 
inter-layer AFM order, is readily a solution on the GGA level. Starting therefrom, DFT+DMFT 
quickly converges towards the same-kind many-body A-AFM phase at low temperatures 
(see Fig.~\ref{fig3:pg}a,b). Note that this is not a strict bulk-like A-AFM ordering, but the
opposite Ti1 layers sandwiching SrO have identical FM direction with comparatively small
magnetic moment. In addition, both Ti5 layers at the respective cell boundary are also FM 
aligned.

There is strongly reduced total spectral weight at the Fermi level compared to GGA, but the 
first layer exhibits spin-polarized QP-like peaks at $\varepsilon_{\rm F}$ associated with an 
Fermi-liquid-like exponent $\alpha=0.95$. The delicate second FM layer is again strongly orbital- 
as well as spin-polarized, and notably already insulating.
Intra-layer (or G-type) AFM ordering is not a strong competitor, although various
starting points and mixing schemes were applied to stabilize such a metastable
solution. Yet the introduced hole doping should indeed weaken the effective 
(because of strong orbital polarization) half-filled strong-AFM scenario in favor of FM 
tendencies. Thus part-FM order, especially close to the doping layer and for SmTiO$_3$, is not 
that surprising. 

An important observation is made, which delivers information concerning 
the many-body fluctuations: when starting from the previous PM solution and allowing for 
spin polarization, DFT+DMFT converges back to the original PM phase via an intriuging 
intermediate pseudogap state (cf. Fig~\ref{fig3:pg}c). The prominent pseudogap signature 
$\Delta_{\rm PG}\sim$0.1eV appears after about 15 self-consistency stemps and is quasi-stable 
for many further calculational steps. Therein, local moments are rather small, i.e. 
$m(\mbox{Ti1-5})=(0.0, -0.02, 0.01, 0.09, 0.04)\,\mu_{\rm B}$.
Interestingly, the pseudogap is associated with orbital- and spin-balanced
electron characteristics in the first layer. The parameters $C_0$ and $\alpha$
are still Fermi-liquid-like in that layer. Yet for Ti2 especially the 
intercept is rather large with $C_0\sim -0.25\,$eV in the pseudogap state. This underlines 
the strong NFL electron-electron scattering in the second layer.

Hence the electronic states attached to SrO are rather fragile and strongly affected by 
fluctuations around the PM solution. Due to the dominant magnetic instability in 
$\delta$-doped SmTiO$_3$, the pseudogap fingerprint in an intermediate state is interpreted 
to originate from the proximity to the AFM-to-FM instability.
Scattering of first-layer itinerant electrons at emerging moments in the deeper
layers below causes a significant spectral-weight reduction at $\varepsilon_{\rm F}$.
Two older perspectives are worth mentioning in this context. Non-Fermi-liquid behavior has been 
observed in DMFT calculations for a two-orbital Hubbard model due to double-exchange physics
induced by orbital selectivity~\cite{biemed05}. Seemingly there are some similarities with
the present findingsd, though the two-orbital scenario is here replaced by an effective 
{\sl two-layer} scenario. In view of the experimental $T^{5/3}$-law for the resistivity, such an 
exponent is obtained for a metal prone to ferromagnetism~\cite{mat68} and was e.g. discussed in 
the context of Ni$_3$Al~\cite{nik05}. This may underline the relevance of additional FM 
fluctuations in the present AFM-based system.

An emerging pseudogap in few-layer SrO-doped SmTiO$_3$ is indeed reported in  
recent tunneling-spectroscopy experiments~\cite{ste16}. Pseudogap behavior is well known for 
the twodimensional one-band Hubbard model proximate to 
AFM order~\cite{dei96,hus01,kyu06,rub09,kat09}, described beyond single-site DMFT. 
But here, intriguing intra/inter-layer-resolved self-energies are sufficient to provide 
a multi-orbital fingerprint of such a fluctuation-dominated phase in the pre-converged 
DFT+DMFT cycle. 
Full stabilization of a pseudogap phase would ask for inter-site self-energies in the 
theoretical description. However note that the present pseudogap fingerprint does not emerge 
from sole in-plane correlations {\sl parallel} to the interface, but additionally from 
perpendicular-to-interface correlations. This may create room for novel designing 
options of this fluctuation physics in terms of different layering/spacing.

To summarize, we find layer-dependent multi-orbital metal-insulator transitions
in $\delta$-doped SmTiO$_3$ with delicate temperature dependence. Unconventional metallicity 
for the two TiO$_2$ layers close to the SrO doping layer is revealed and 
nearly completely orbital-polarized Mott-insulating layers beyond. The next-nearest
TiO$_2$ layer is critical in the sense that it mediates between the Fermi-liquid
layer and the Mott layers, resulting in NFL behavior in line with experimental findings. 
This NFL regime is associated with strong AFM-to-FM spin fluctuations that may lead to a 
pseudogap structure at the Fermi level. 
Finally, the present many-body oxide-heterostructure study shall stimulate the investigation 
of novel emergent electronic phases~\cite{wan12,cao16} by advanced theoretical means
on the realistic level.

\begin{acknowledgments}
The author is indebted to S. Stemmer for helpful discussions.
This research was supported by the Deutsche Forschungsgemeinschaft through FOR1346. 
Computations were performed at the University of Hamburg and the JURECA 
Cluster of the J\"ulich Supercomputing Centre (JSC) under project number hhh08.
\end{acknowledgments}

\bibliographystyle{apsrev}
\bibliography{bibextra}

\newpage

\begin{center}
{\LARGE Supplemental Information}
\end{center}

\section{DFT+DMFT approach}
The charge self-consistent DFT+DMFT method is employed. For the DFT part, a mixed-basis 
pseudopotential method~\cite{lou79,mbpp_code} build on norm-conserving pseudopotentials
as well as a combined basis of localized functions and plane waves. The generalized-gradient 
approximation (GGA) in the Perdew-Burke-Ernzerhof form~\cite{per96}, is put into practice.
The partially-filled Sm($4f$) shell is treated in the frozen-core approximation since the
highly-localized $4f$ electrons do not have key influence on the present doped-Mott
physics. In the mixed-basis, localized functions are included for Ti($3d$) as well as
for O($2s$), O($2p$) to reduce the energy cutoff $E_{\rm cut}$ for the plane waves.

Our correlated subspace consists of the effective Ti($t_{2g}$) Wannier-like functions 
$w_n(t_{2g})$, i.e. is locally threefold. The $w(t_{2g})$ functions are obtained from the 
projected-local-orbital formalism~\cite{ama08,ani05,aic09,hau10}, using as projection functions 
the linear combinations of atomic $t_{2g}$ orbitals, diagonalizing the Ti  
$w_n(t_{2g})$-orbital-density matrix. A band manifold of 60 $t_{2g}$-dominated Kohn-Sham 
states at lower energy are used to realize the projection. Local Coulomb interactions in 
Slater-Kanamori form for the $w_n(t_{2g})$ orbitals are parametrized by a 
Hubbard $U=5\,$eV and a Hund's coupling $J_{\rm H}=0.64\,$eV. These values for the local 
Coulomb interactions are common for correlated titanates~\cite{pav04}. 
The single-site DMFT impurity problems are solved by the continuous-time quantum 
Monte Carlo scheme~\cite{rub05,wer06} as implemented in the TRIQS package~\cite{par15,set16}. 
A double-counting correction of fully-localized 
type~\cite{ani93} is utilized. To obtain the spectral information, analytical
continuation from Matsubara space via the maximum-entropy method is performed. About 40-50 
DFT+DMFT iterations (of alternating Kohn-Sham and DMFT impurity steps) are necessary for 
full convergence.

\section{Magnetic order in bulk samarium titanate}
\begin{figure}[t]
\centering
\includegraphics*[width=8.25cm]{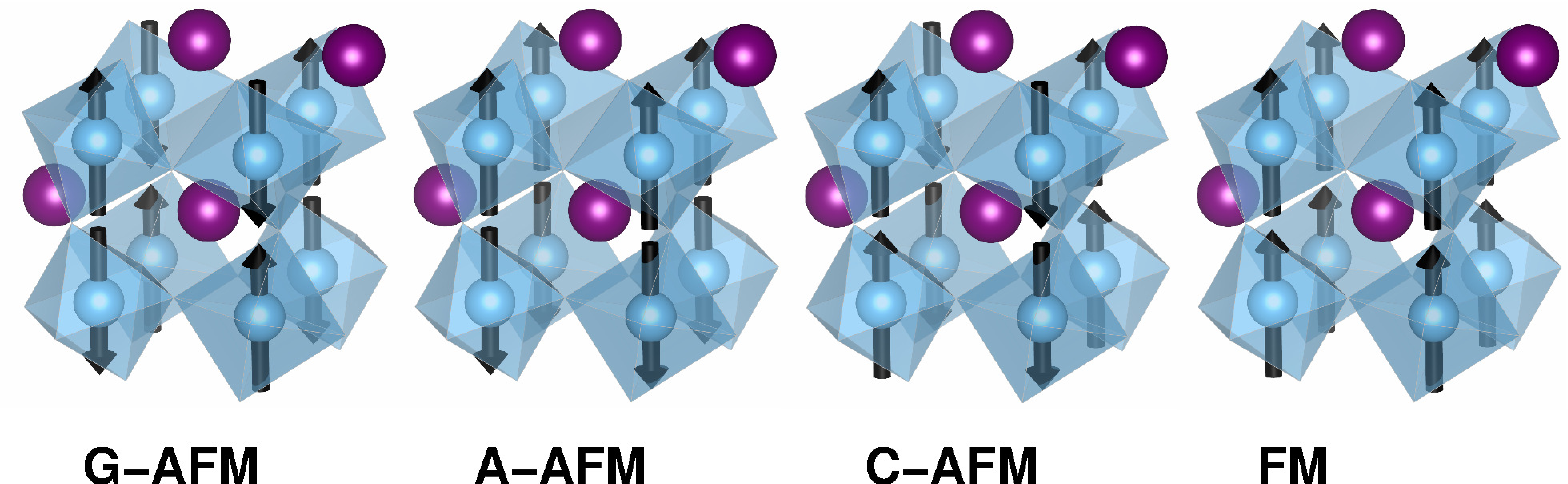}
\caption{Competing magnetic orders in SmTiO$_3$.}\label{fig:orders}
\end{figure}

Bulk SmTiO$_3$ displays antiferromagnetic (AFM) order of G-type form below 
$T_{\rm N}=45\,$K. Yet the compound is just at the border towards ferromagnetism, 
since already the next rare-earth titanate $R$TiO$_3$ ($R$: rare-earth element) 
in the given series, i.e. GdTiO$_3$, is ferromagnetic (FM). To provide insight into 
the subtle competition between different magnetic orders, we compute the total energies
within GGA and GGA+U for the three most prominent AFM orderings, i.e. G-type, A-type and
C-type as well as for FM ordering (see Fig.~\ref{fig:orders}). A plane-wave energy cutoff 
$E_{\rm cut}=16\,$Ryd and a $9\times9\times9$ $k$-point grid is applied.
The calculations are based on the 100\,K experimental crystal data of 
Komarek {\sl et al.}~\cite{kom07} and the resulting energies and Ti magnetic moments
are displayed in Tab.~\ref{tab:orders}.

All ordering cases are at least metastable and the nonmagnetic energy lies always higher, 
i.e. magnetic ordering is strongly favored at low temperatures. Note however that 
DFT(+U), in contrast to DFT+DMFT, does not allow to treat the true {\sl paramagnetic} case.
For each order, GGA(+U) yields metallic(insulating) solutions.
The different ordered states are all very close in energy, at most they differ in about
10(1)\,meV/atom within GGA(+U). Albeit the numbers designate FM order as lowest, given
the DFT approximation, especially for the Mott-insulating designed GGA+U an energy 
difference of 1\,meV/atom is usually beyond the limit of physical signficance.
But the quasi-degeneracy of the magnetic orders is nonethless physically sound, since
as noted,  SmTiO$_3$ is just at the AFM-to-FM transition point in the rare-earth titanate 
series.
\begin{table}[t]
\begin{ruledtabular}
\begin{tabular}{l|rrrr}
     scheme  & G-AFM  & A-AFM  & C-AFM  & FM \\ \hline
             & 11.3   & 2.1    & 10.8   & 0.0 \\
\bw{GGA}     & 0.32   & 0.62   & 0.46   & 0.72 \\[0.1cm]
             &  1.2   & 0.3    &  1.2   & 0.0 \\
\bw{GGA+U}   & 0.73   & 0.75   & 0.74   & 0.76 \\
\end{tabular}
\end{ruledtabular}
\caption{Energy and magnetic moment for different magnetic orders
in SmTiO$_3$. 
First line: energy $E_{\rm mag}$ (in meV/atom) of the magnetically 
ordered state with respect to the lowest energy in the given set.
Second line: local Ti magnetic moment (in $\mu_{\rm B}$). 
For GGA+U, the values $U=5\,$eV and $J_{\rm H}=0.64\,$eV are 
chosen.}\label{tab:orders}
\end{table}

\section{Structural Details for the $\delta$-doped architecture}
The $\delta$-doped SmTiO$_3$ architecture is realized by a superlattice based on a 100-atom 
unit-cell. It consists of 10(9) TiO$_2$(SmO) layers, separated by a single SrO monolayer. 
Each TiO$_2$ layer is build from two possibly symmetry-inequivalent Ti ions, to allow for 
potential intra-layer spin or charge ordering. Together with the five-layer resolution away 
from the doping layer, there are thus 10 inequivalent Ti single-site DMFT problems in our 
realistic modelling. 
The original lattice parameters~\cite{kom07} are brought in the directional 
form of the experimental works Ref.~\cite{jac14,mik15}, but without lowering the $Pbnm$ 
symmetry. The original $c$-axis 
is {\sl parallel} to the doping layer and the original $a,b$-axes are respectively inclined. 
The plane-wave energy cutoff is set to $E_{\rm cut}=11\,$Ryd and a $5\times5\times3$ $k$-point 
grid is used.
With fixed lattice parameters, all atomic positions in the supercell are structurally 
relaxed within DFT(GGA) until the maximum individual atomic force settles below 5\,mRyd/a.u.. 
The lattice distortion introduced by the SrO layer is well captured by the structural 
relaxations. No relaxation of the lattice paramteres is performed. A change of lattice 
parameters is  very small due to the structural similarity and does not invoke changes of 
the key physics discussed in the given work. The obtained crystal structure is used for the 
PM, A-AFM as well as the pre-converged electronic structure studies.

\section{Influence of  the local-interaction strength}
\begin{figure}[b]
\centering
\includegraphics*[width=8cm]{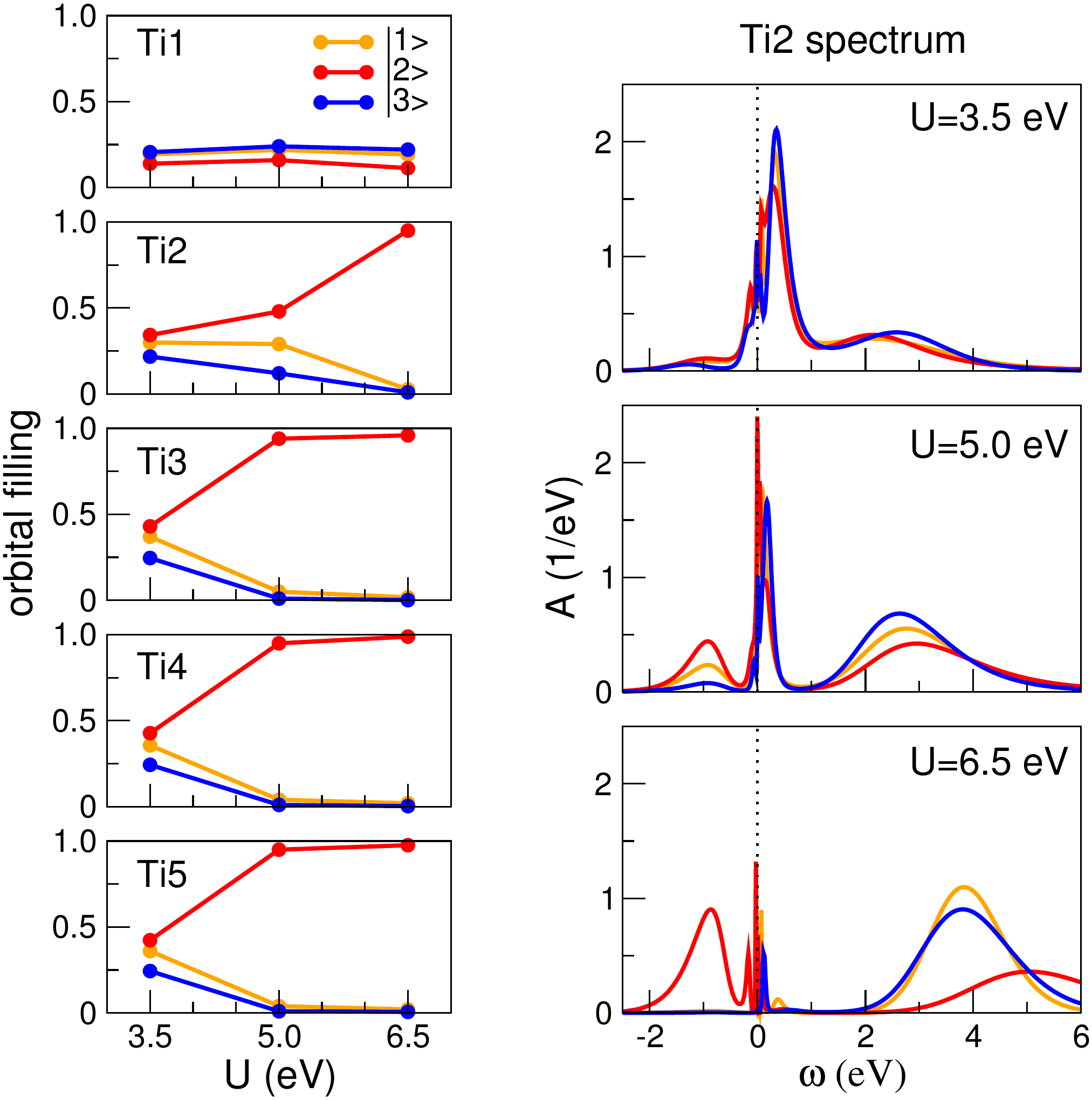}
\caption{(color online) Local Ti data for different values of Hubbard $U$ and
fixed $J_{\rm H}=0.64\,$eV ($T=48$K). Left: layer-dependent paramagnetic orbital 
filling (with no difference between intra-layer Ti:1,Ti:2). 
Right: local Ti2 spectral function with respect to $U$.}\label{fig2:udep}
\end{figure}
The chosen local Coulomb interactions are well established for bulk titanates~\cite{pav04}. 
A first-principles computation of these parameters for the large $\delta$-doped architecture,  
including their layer dependence, is currently numerically unfeasible. In order to still
examine the influence of smaller/larger local Coulomb interaction, we additionally studied
the DFT+DMFT electronic structure for $U=3.5\,$eV and $U=6.5\,$eV. The Hund's exchange is
even less sensitive and therefore remains fixed at $J_{\rm H}=0.64\,$eV.

Figure~\ref{fig2:udep} exhibits the layer-dependent Ti orbital filling with $U$. As expected,
for the smaller value of $U$, the orbital polarization is much weaker for the layers beyond Ti1.
Also the Mott state is not reached with distance from the doping layer up to Ti5, thus 
$U=3.5\,$eV appears too small to account for the correct $\delta$-doping physics. On the
other hand, for $U=6.5\,$eV the orbital polarization for Ti3-5 is even increased. Moreover,
the second TiO$_2$ layer with Ti2 is also strongly orbital polarized for this larger $U$. 
The additionally shown Ti2 local spectral function in Fig.~\ref{fig2:udep} reveals that 
the second layer is about to enter a Mott-insulating state. Hence a substantially larger, 
somewhat unphysical, Hubbard $U$ destroys the two-layer dichotomy. 

\section{Details on the self-energy fitting}
The low-frequency characteristics of the local electronic self-energy within the two Ti layers 
closest to the SrO doping layer, i.e. Ti1 and Ti2, is analyzed in some detail in the main text. 
This allows to extract important information on the (non)-Fermi-liquid quality connected to
states close to the Fermi energy. To asses this quality, we focus on two features of the
self-energy, namely the intercept $C_0=\lim_{\omega\rightarrow 0}\Sigma$ and the form 
$\tilde{Z}(\omega)=
\left(1-\left.\frac{\partial\Sigma}{\partial\omega}\right|_{\omega\rightarrow 0}\right)^{-1}$.
For a Fermi liquid (FL) at low temperature, the intercept $C_0$ approaches zero and 
$\tilde{Z}(\omega)=Z$ is the constant quasiparticle (QP) weight. Significant deviations from 
these features signals a non-Fermi liquid (NFL).

We here provide further information on the self-energy fitting procedure in the $\delta$-doped
paramagnetic phase to enable such a discrimination, as well as on the grade of those fits.
The relevant object in this context is the imaginary part of the Matsubara self-energy. In correct 
mathematical terms, this function reads ${\rm Im}\,\Sigma(i\omega_n)$, with the fermionic 
Matsubara frequencies defined as $\omega_n=(2n+1)\pi\,T$. From a sole fitting-function point of 
view, we however refer in the following to ${\rm Im}\,\Sigma(\omega_n)$. Moreover we discuss
the fitting parameters as dimensionless, imagining proper normalization.

\begin{figure}[t]
\centering
\includegraphics*[width=8.25cm]{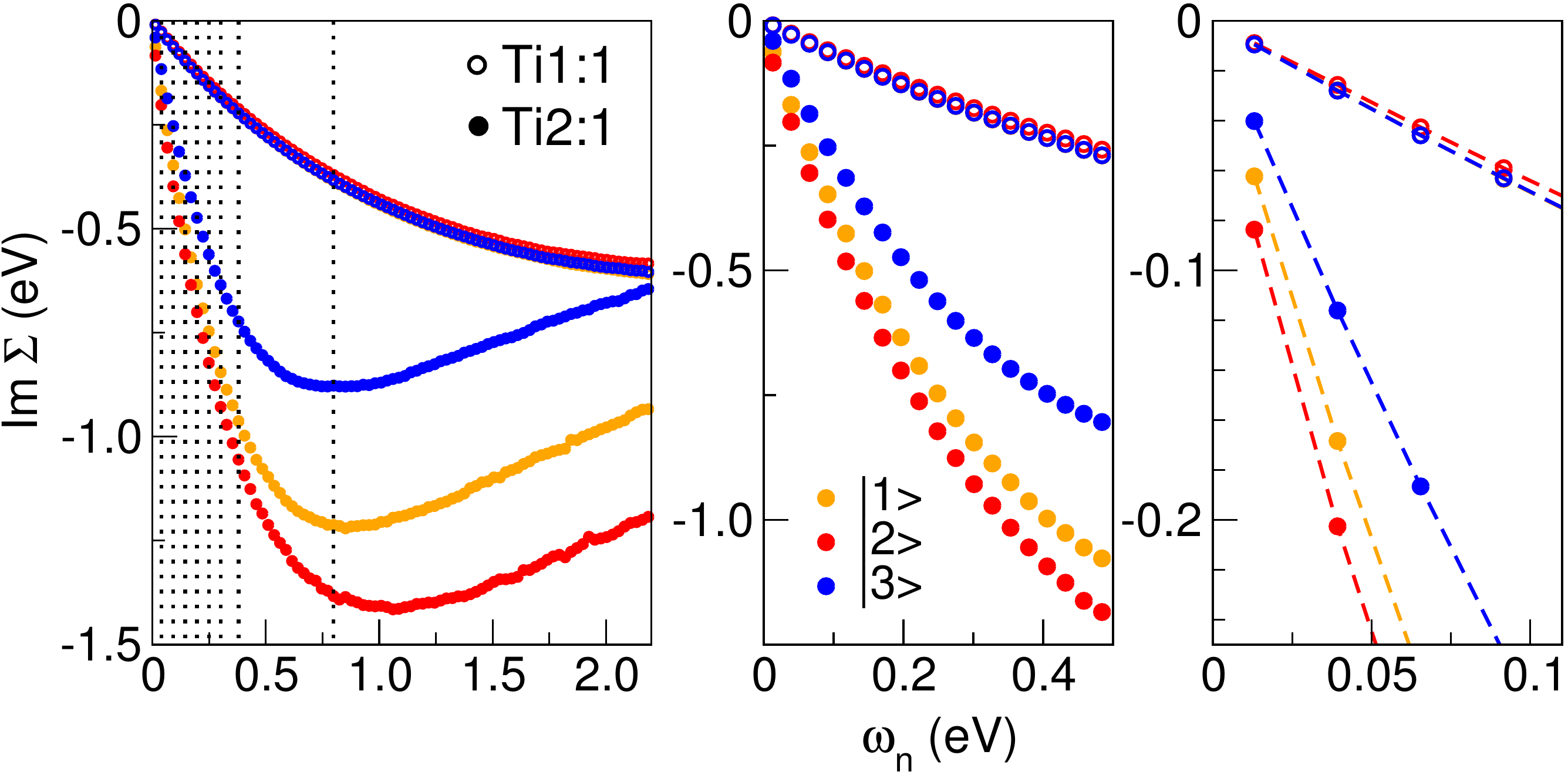}
\caption{(color online) Imaginary part of the Ti self-energy on the Matsubara axis for the
first two TiO$_2$ layers with different resolution from left to right.
Left: dotted lines mark the frequency cutoff $n_c=2,4,6,8,10,12,16,32$ in the different 
fits (see Tab.~\ref{tab:fit}). Dashed lines in right part are no fitting curves but guides 
to the eyes.}\label{fig:fit}
\end{figure}

Figure~\ref{fig:fit} shows different resolutions of the imaginary part of the multi-orbital 
self-energy with respect to the frequencies $\omega_n$. The quantum-Monte-Carlo data is well 
converged at $T=48\,$K to facilitate our low-frequency examination. In order to check the 
influence on the frequency cutoff $n_c$ (i.e. all frequencies with $n\le n_c$ are used for the
fit), different values for $n_c$ are chosen. Two different stages of fitting procdures are 
performed and results are here exemplified for the dominant Ti $|2\rangle$ state in the
first (Ti1) and second (Ti2) layer next to SrO (see Tab.~\ref{tab:fit}).
\begin{table}[t]
\begin{ruledtabular}
\begin{tabular}{r|cc|ccc}
             & \multicolumn{2}{c}{$\displaystyle{\rm Im}\,\Sigma(\omega_n)=\sum_{k=0}^{1,5}a_k\,\omega_n^k$} & \multicolumn{3}{c}{$\displaystyle{\rm Im}\,\Sigma(\omega_n)=C_0+A\,\omega_n^{\alpha}$} \\[0.4cm]
$n_c$   & $a_0=C_0$      & $(1-a_1)^{-1}$=$Z$ & $C_0$    & $A$  & $\alpha$ \\ \hline
2            & -0.0006    & 0.61 & $-$        & $-$   & $-$ \\
4            & -0.0006    & 0.61 & $<10^{-4}$ & -0.61 & 0.98 \\
6            & -0.0016    & 0.66 & $<10^{-4}$ & -0.58 & 0.96 \\
8            & -0.0009    & 0.63 & $<10^{-4}$ & -0.56 & 0.94 \\
10           & -0.0009    & 0.63 & $<10^{-4}$ & -0.54 & 0.93 \\
12           & -0.0004    & 0.61 & $<10^{-4}$ & -0.53 & 0.92 \\
16           & $<10^{-4}$ & 0.60 & $<10^{-4}$ & -0.51 & 0.90  \\
32           & $<10^{-4}$ & 0.60 & 0.0002     & -0.62 & 0.98 \\[0.2cm] \hline
2            & -0.0243    & 0.18 & $-$        & $-$   & $-$ \\
4            & -0.0382    & 0.20 & -0.0131    & -2.96 & 0.85 \\
6            & -0.0111    & 0.14 & -0.0111    & -2.68 & 0.81 \\
8            & -0.0171    & 0.16 & -0.0084    & -2.50 & 0.78 \\
10           & -0.0149    & 0.15 & -0.0046    & -2.36 & 0.75 \\
12           & -0.0177    & 0.16 & $<10^{-4}$ & -2.25 & 0.72 \\
16           & -0.0189    & 0.16 & $<10^{-4}$ & -2.11 & 0.69 \\
32           & -0.0267    & 0.17 & $<10^{-4}$ & -2.69 & 0.80 \\
\end{tabular}
\end{ruledtabular}
\caption{Detailed results for the least-mean-squares fitting procedures applied to 
${\rm Im}\,\Sigma(\omega_n)$ in the $\delta$-doped case, shown for the Ti 
$|2\rangle$ state in the first (top) and the second TiO$_2$ layer (bottom) next 
to the SrO doping layer. Left columns: polynomial fit, of first order for $n_c=2-4$ and
of fifth order for $n_c=6-32$. Right colums: exponential fit for $n_c\ge 4$.}\label{tab:fit}
\end{table}
First, motivated by {\sl assuming} Fermi-liquid theory holds, a polynomial fit of simplistic 
first order ($n_c\le 4$) and of 5th order ($n_c\ge 6$) is processed. Intercept and $Z$ are 
therefrom easily extracted from the zeroth and first-order terms, respectively. A good quality of
the fitting shall focus on the very low-frequency region, but takes care of the $\omega_n$
evolution. Hence the sole linear fitting for very small $n_c$ as well as the higher-order 
fit for rather large $n_c\sim 16-32$ are less suited to the problem. But the results show that
a proper fitting leads to robust values for $C_0$ and $Z$ with e.g. an error $\Delta Z\sim 0.03$.

Questioning Fermi-liquid behavior asks for an exponential fitting function with an exponent 
$\alpha$ (cf. Tab.~\ref{tab:fit}). Only for $\alpha=1$ a constant QP weight and thus
Fermi-liquid behavior is recovered. However the fitting is more exclusively restricted to small
$n_c$. A large $n_c$ results in a too small intercept $C_0$, since the fitting function tries
to account for the natural bending of ${\rm Im}\,\Sigma(\omega_n)$ at larger $\omega_n$ by
shifting $C_0$ towards zero. Still, the quality of our numerical 
data is high enough to reveal the resilient FL-to-NFL crossover from the first to the second
TiO$_2$ layer. In the second layer, the exponent $\alpha$ manifestly deviates from unity and the
intercept $C_0$ from zero. Hence electron-electron scattering is beyond FL theory and does not
easily vanish at the Fermi energy. Of course the previous FL statements only fully hold for
$T\rightarrow 0$, but our temperature is already well below a possible QP coherency scale and
the contrasting behavior in the seemingly FL-like first TiO$_2$ layer is obvious.

\end{document}